\documentclass{article}
\usepackage{amsmath}
\usepackage{graphicx}
\usepackage{xcolor}
\usepackage{appendix}
\usepackage{latexsym, amsfonts, amssymb, txfonts, pxfonts}
\usepackage{jheppubbu}

\title{Ladder operators of the de Sitter algebra}            
\author[a]{Manizheh Botshekananfard}
\author[a]{Elif Büşra Güraksın}
\author[a,b]{Gizem \c{S}eng\"or}
\affiliation[a]{Physics Department, Boğaziçi University\\
34342 Bebek, İstanbul Turkey}
\affiliation[b]{Feza Gürsey Center for Physics and Mathematics, Boğaziçi University, Kandilli Kampüs, Feza Gürsey Binası,\\ Kandilli Mah., Rasathane Cad., 34684, İstanbul, Türkiye}
\affiliation[b]{}
\emailAdd{gizem.sengor@bogazici.edu.tr}
\emailAdd{}
\abstract{We identify raising and lowering operators of the de Sitter algebra with
focus on their action on states in particular in 4 spacetime
dimensions. There isn't a unique solution to the question of how the de
Sitter ladder operators act on states. By fixing the action of certain
generators one can conclude on the action of the rest. Our main aim is
to be able to identify highest and lowest weight states such that we can
recognize them for quantized fields on a rigid de Sitter background.}
\begin{document}

\maketitle

\newpage

\section{Introduction}
\label{sec:Intro} 

Our main question in this work is: if we can construct a tower of states starting with a lowest weight state of the de Siter algebra $\mathfrak{so}(d+1,1)$, under what conditions can we obtain finite and infinite towers? To answer this question, we first have to understand what is the weight that can be raised and lowered, and what defines the lowest weight state. Our goal is to be able to construct towers of states where each step corresponds to a unitary irreducible representation (UIR) of the de Sitter group $SO(d+1,1)$.

In section \ref{sec:Identifying raising and lowering} we will recognize that translation and special conformal transformation (SCT) generators take the role of raising and lowering operators by raising and lowering the value of the scaling dimension by one respectively. Hence the scaling dimension is the weight we can raise and lower to construct the towers. From the point of representation theory only discrete series representations correspond to UIRs with integer values of scaling dimensions and hence the towers of UIRs will only be made up of discrete series representations. We will fix the action of the SCT generators such that the UIR state corresponding to the initial step of a tower is the lowest weight state. The de Sitter algebra allows for some freedom in fixing the action of the Translation generator which we will discuss in section \ref{sec:Fixing the action of Raising and Lowering operators}. 

To construct our towers, we will pick the lowest weight state from among the discrete series representations of the group $SO(4,1)$ which corresponds to the isometry group of four dimensional de Sitter spacetime ($dS_4$) and the conformal group of three dimensional Euclidean space. We are interested in discrete series representations realized in terms of symmetric traceless tensors (which are only accessible when $d$ equals to $1$ and $3$ \cite{Dobrev:1977qv}) and which arise in the context of free quantum fields on a fixed de Sitter background. We will base our discussion on the representations that the free massless scalar on $dS_4$ give rise to at the late-time boundary of $dS_4$, as discussed in \cite{sengor2021scalar}. Labeling the de Sitter group representations by the scaling dimension $\Delta$, the dilatation eigenvalue, and spin $l$, related to the $\mathfrak{so}(3)$ Casimir eigenvalue, as $|\Delta,l\rangle$, the free massless scalar on $dS_4$ gives rise to two states labeled by $|0,0\rangle$ and $|3,0\rangle$. Strictly speaking these representations correspond to exceptional series type-I unitary irreducible representations while the category discrete series arise under exceptional series type-II, which start at nonzero spin. Yet focusing on  the massless scalar representations will be enough for our purposes.

In section \ref{sec:Lowest weight towers of a massless scalar} we will use the results of previous sections to construct two towers out of the discrete series states that arise at the late-time boundary of $dS_4$ from a free  massless scalar field.  Our construction will create a single step tower out of the lowest weight state with $|\Delta,l\rangle=|0,0\rangle$ while the lowest weight state with $|3,0\rangle$ will give rise to an infinite tower that looks more like a ladder with different values of scaling dimension and spin at every step. 

\section{Identifying the Raising and Lowering operators}
\label{sec:Identifying raising and lowering}
Following the conventions of \cite{Dobrev:1977qv}, the $so(d+1,1)$ algebra with its generators being dilatation ($D$), translations ($T_i$), SCT ($C_i$) and $SO(d)$ rotations ($M_{ij}$), is as follows
\begin{subequations}
\label{full algebra}
    \begin{align}
    \label{full algebra1}    [D,C_i]&=-C_i,~~[D,T_i]=T_i,~~[T_i,C_j]=2D\delta_{ij}-2M_{ij}\\
        \label{rot tr}[M_{ij},T_k]&=\delta_{ik}T_j-\delta_{jk}T_i,\\
        \label{rot sct}[M_{ij},C_k]&=\delta_{ik}C_j-\delta_{jk}C_i.
    \end{align}
\end{subequations}
The quadratic casimir eigenvalue is 
\begin{equation}
\label{eqn:so(4,1)casimirdeltaemph}c_{so(d+1,1)}=\Delta(\Delta-d)+l(l+d-2).
\end{equation}
We focus on dimensions with $d\geq 3$ and label states by the eigenvalue of the $so(d)$ quadratic Casimir,\footnote{To be explicit, in our notation we are following \cite{Dobrev:1977qv}, where Casimir operators are defined by the sum $\mathcal{C}=\frac{1}{2}L_{AB}L^{BA}$. This has an overall minus in the definition compared to \cite{sun2021note,Silva_note}. For the $so(d)$ Casimir we write  $\mathcal{C}_{so(d)}=\frac{1}{2}M_{ij}M^{ji}$.} and the Dilatation generator
   \begin{subequations}
       \begin{align}
    \label{eqn:M2onstate}       \mathcal{C}_{so(d)}\Big|\Delta,l\Big\rangle&=l(l+d-2)\Big|\Delta,l\Big\rangle,\\
 D|\Delta,l\Big\rangle&=\Delta |\Delta,l\Big\rangle.
       \end{align}
   \end{subequations}
 
As also pointed out in \cite{sun2021note}, the commutations between dilatation and SCT generator in equation \eqref{full algebra1} implies that the special conformal transformation generators lower the dilatation eigenvalue by one, while translation generators raise it. To see this, consider acting on $|\Delta,l\rangle$ with these commutation relations. Commutation between dilatation and SCT gives
\begin{subequations}
    \begin{align}
        \left[D,C_i\right]|\Delta,l\rangle&=-C_i|\Delta,l\rangle\\
        D\left(C_i|\Delta,l\rangle\right)&=(\Delta-1)\left(C_i|\Delta,l\rangle\right).
    \end{align}
\end{subequations}
This means that the state $C_i|\Delta,l\rangle$ is an eigenstate of the dilatation operator with scaling dimension $\Delta-1$ and the special conformal transformation generators act as the lowering operators of the $so(d+1,1)$ algebra.

 Similarly , the commutation of the dilatation and translation generators give
\begin{align}
D\left(T_i|\Delta,l\rangle\right)&=\left(\Delta+1\right)\left(T_i|\Delta,l\rangle\right),
\end{align}
which implies that the translation generators act as the raising operators of the $so(d+1,1)$ algebra.

Now that we have identified the translation generators $T_i$ as the operators that raise the Dilatation eigenvalue $\Delta$ and the special conformal generators $C_i$ as the operators that lower it, lets look at how these generators act on the label $l$ associated with spin. Since $T_i$ and $C_i$ are part of the algebra, they will not change the $\mathfrak{so}(4,1)$ Casimir eigenvalue\footnote{ We thank Vasileios Letisos and Joris Raeymaekers for helpful discussions on this point.}. This means they must change the spin value. Let us label the new states as
\begin{subequations}
    \begin{align}
        \label{newC}C_i|\Delta,l\rangle &\equiv|\Delta_C,l_C\rangle,~~\text{where}~~\Delta_C=\Delta-1\\
T_i|\Delta,l\rangle &\equiv|\Delta_T,l_T\rangle,~~\text{where}~~\Delta_T=\Delta+1.
    \end{align}
\end{subequations}
To determine how the spin value $l$ changes, we will demand that the new states also be eigenstates of the $\mathfrak{so}(4,1)$ casimir  corresponding to the same eigenvalue as the original state $|\Delta,l\rangle$. That is we demand that
\begin{subequations}
    \begin{align}
        \Delta(\Delta-d)+l(l+d-2)&=\Delta_C(\Delta_C-d)+l_C(l_C+d-2),\\
        \Delta(\Delta-d)+l(l+d-2)&=\Delta_T(\Delta_T-d)+l_T(l_T+d-2)
    \end{align}
\end{subequations}
under the action of SCT and translations respectively.\footnote{The commutation between rotation and translation generators as given in \eqref{rot tr} imply the following commutation relation between $so(d)$ quadratic casimir and translation generator
\begin{align}
 \label{so3 T comm}  \left[\mathcal{C}_{so(d)},T_k\right]=M_{ik}T^i+T^iM_{ik}.
\end{align}
By \eqref{rot sct}, the commutation relation between $
\mathcal{C}_{so(d)}$ and the special conformal transformation generators will be of the same form. However these relations are not practical in figuring out how the spin label changes.} These equations are solved by
\begin{subequations}
\label{new spin values}
    \begin{align}
 \label{new spin C}       l_C&=1-\frac{d}{2}\pm\sqrt{-2d+\frac{d^2}{4}-2l+dl+l^2+2\Delta},\\
 \label{new spin T}       l_T&=1-\frac{d}{2}\pm\sqrt{\frac{d^2}{4}-2l+dl+l^2-2\Delta}.        
    \end{align}
\end{subequations}

Up to constants $\mathcal{N}_C$ and $\mathcal{N}_T$ and the spin values $l_C$, $l_T$ to be determined, we can identify
\begin{subequations}
\label{C T undet}
\begin{align}
 \label{eqn:so(4,1)lowering}
C_i|\Delta,l\rangle &\equiv\mathcal{N}_C(\Delta,l)|\Delta-1,l_C\rangle,\\
\label{eqn:so(4,1)raising}T_i|\Delta,l\rangle &\equiv\mathcal{N}_T(\Delta,l)|\Delta+1,l_T\rangle.
\end{align}
\end{subequations}
We will discuss how to obtain meaningful values of $l_C$ and $l_T$ next. At this level, from among the algebra \eqref{full algebra}, equations \eqref{C T undet} imply that the action of $[T_i,C_i]$ on a state $|\Delta,l\rangle$  requires
\begin{align}
    \label{eqn:NCT dilatation recursion0}\mathcal{N}_ C(\Delta,l)\mathcal{N}_ T(\Delta-1,l_C)-\mathcal{N}_ T(\Delta,l)\mathcal{N}_ C(\Delta+1,l_T)&=2\Delta
\end{align}
with 
\begin{align}\label{raise and lower}
C_i|\Delta+1,l_T\rangle=\mathcal{N}_C(\Delta+1,l_T)|\Delta,l\rangle,~~T_i|\Delta-1,l_C\rangle=\mathcal{N}_T(\Delta-1,l_C)|\Delta,l\rangle.
\end{align}
In other words implying that after raising and lowering back to back one must end up with the initial state up to some overall constant. The $\mathfrak{so}(d+1,1)$ quadratic Casimir can be written in terms of the generators as
\begin{align}
   \mathcal{C}_{\mathfrak{so}(d+1,1)}=\mathcal{C}_{\mathfrak{so}(d)}+D^2+d D+\sum^d_{i=1}C_iT_i. 
\end{align}
 Together with our conventions in \eqref{C T undet} and \eqref{raise and lower}, the quadratic Casimir eigenvalue equation \eqref{eqn:so(4,1)casimirdeltaemph} implies
 \begin{equation}
    \label{eqn: NCT casimir recursion0} \mathcal{N}_T(\Delta,l)\mathcal{N}_C(\Delta+1,l_T)=-2\Delta.  
 \end{equation}

 \section{Fixing the action of Raising and Lowering operators}
 \label{sec:Fixing the action of Raising and Lowering operators}
In the previous section we fixed the action of the raising and lowering generators up to undetermined constants $\mathcal{N}_C$ and $\mathcal{N}_T$, which are governed by the following equations
\begin{subequations}\label{eqn:recursions}
    \begin{align}
        \label{eqn:NCT dilatation recursion}\mathcal{N}_ C(\Delta,l)\mathcal{N}_ T(\Delta-1,l_C)-\mathcal{N}_ T(\Delta,l)\mathcal{N}_ C(\Delta+1,l_T)&=2\Delta,\\
        \label{eqn: NCT casimir recursion} \mathcal{N}_T(\Delta,l)\mathcal{N}_C(\Delta+1,l_T)&=-2\Delta.
    \end{align}
\end{subequations}
Notice that the two equations together imply
\begin{align}
    \label{eqn: NCT casimir into recursion consistency}
    \mathcal{N}_ C(\Delta,l)\mathcal{N}_ T(\Delta-1,l_C)&=0.
\end{align}
One does not have an exact solution to these equations. Instead these equations can be solved either by fixing the action of the lowering operator $C_i$ or the raising operator $T_i$ for a given $\Delta$ at a time. Equation \eqref{eqn: NCT casimir into recursion consistency} hints two possibilities along this line:
\begin{subequations}
\label{eqn:cases}
    \begin{align}
        \text{case 1:}& ~~\mathcal{N}_ C(\Delta,l)=0,\\
        \text{case 2:}& ~~\mathcal{N}_ T(\Delta-1,l_C)=0.
    \end{align}
\end{subequations}
Since motivation was to build lowest weight towers from a given state, we go along with case 1 and demand the starting state of interest $|\Delta,l\rangle$ to be annihilated by the lowering operator $C_i$. That is we demand
\begin{align}
\mathcal{N}_C(\Delta_s,l_s)=0~~\text{for a specified lowest weight state}~~|\Delta_s,l_s\rangle.
\end{align}
This eliminates the rather likely troublesome possibility that we may end up with complex values of spin under the action of $C_i$ by equation \eqref{new spin C}.

Setting $\mathcal{N}_ C(\Delta_s,l_s)=0$ for a particular $\Delta_s$ leaves the action of $\mathcal{N}_ T(\Delta_s,l_s)$ free. Then for a fixed $\mathcal{N}_ T(\Delta_s,l_s)$ one has $\mathcal{N}_ C(\Delta_s+1,l_T)$ determined by equation \eqref{eqn:NCT dilatation recursion} as follows
\begin{align}
   \mathcal{N}_ C(\Delta_s+1,l_T)&=-\frac{2\Delta_s}{\mathcal{N}_ T(\Delta_s,l_s)}~~\text{for a given $\Delta_s$ and $l_s$}.
\end{align}
This way each state $|\Delta_s,l_s\rangle$, where the subscript $s$ implies specific choices, is a lowest weight state since $\mathcal{N}_ C(\Delta_s,l_s)=0$ but it can be raised to the state $|\Delta_s+1,l_T\rangle$ and there is a well defined way to lower the state $|\Delta_s+1,l_T\rangle$ back to $|\Delta_s,l_s\rangle$ by the action of $C_i$.

A simple enough yet general choice for the action of $T_i$ would be to take $\mathcal{N}_ T(\Delta,l)$ linear in $\Delta$ and $l$, such as
\begin{align}
   \mathcal{N}_ T(\Delta,l)&=a \Delta+b l  ~~\text{for any $\Delta$ and $l$},
\end{align}
where $a,b$ and $c$ are constants. Then accordingly
\begin{equation}
 \label{NC D+1}   \mathcal{N}_ C(\Delta_s+1,l_T)=-\frac{2\Delta_s}{a \Delta_s+b l_s },
\end{equation}
 and the action of the ladder operators on the lowest weight state $\{\Delta_s,l_s\}$ is
\begin{subequations}\label{case1_choiceab_action}
    \begin{align}
        &C_i|\Delta_s+1,l_T\rangle=-\frac{2\Delta_s}{a\Delta_s+b l}|\Delta_s,l_s\rangle,\\
     |\Delta_s,l_s\rangle:~~&   T_i|\Delta_s,l_s\rangle=\left(a\Delta_s+b l_s\right)|\Delta_s+1,l_T\rangle,~~C_i|\Delta_s,l_s\rangle=0
    \end{align}
\end{subequations}

 \section{Lowest weight towers of a massless scalar}
 \label{sec:Lowest weight towers of a massless scalar}
 
Starting with a state $|\Delta,l\rangle$, that corresponds to a unitary irreducible representation (UIR) of the de Sitter group, we want to see if we can end up with a state that corresponds to another UIR of the de Sitter group under the action of $T_i$ and $C_i$. In the previous section, we saw that the translation and SCT generators change the dilatation eigenvalue by one. From the context of de Sitter representation theory, this implies focusing on discrete series representations, who are the only ones with integer scaling dimension from among the UIRs. Since the discrete series representations, realized by symmetric traceless tensors exist only for $d=\{1,3\}$, in the remainder of the discussion we will focus on $d=3$. 

As we noted in the introduction let us focus on the massless scalar on $dS_4$, which is expected to carry two unitary representations which can be explicitly recognized at the late-time boundary, as the following states $|\Delta,l\rangle$ 
\begin{align}
\label{eqn:realizable exs}    
\text{massless scalar on $dS_4$:}~~|0,0\rangle,~|3,0\rangle,\end{align}
 normalized with respect to the discrete series inner product \cite{sengor2021scalar}. 
 One also expects to be able to find discrete series states with
 \begin{align}
     \text{massless vector on $dS_4$:}~~|1,1\rangle,~|2,1\rangle
\end{align}
from the late-time behavior of a massless vector field on $dS_4$ \cite{sun2021note}.

For $l=0$ in $d=3$, the solution for $l_T$ with the upper sign sets $l_T=1$, which is a physically sensible value for spin.

The choice \eqref{case1_choiceab_action} automatically makes the lowest weight state $|\Delta_s,l_s\rangle=|0,0\rangle$, also be a highest weight state,
\begin{align}\label{case100} |0,0\rangle:~~&T_i|0,0\rangle=0,~~C_i|0,0\rangle=0,  \end{align}
with $\mathcal{N}_C(0,0)=\mathcal{N}_T(0,0)=0$. By \eqref{NC D+1}, $\mathcal{N}_ C(1,1)$ is indeterminate. This is acceptable, since one cannot raise $|0,0\rangle$ up to $|1,1\rangle$, it is not expected to be able to lower $|1,1\rangle$ down to $|0,0\rangle$. Thus being both the lowest and the highest state $|0,0\rangle$ constitutes a ladder of a single step. This is also good that one will not reach the massless vector state from the massless scalar state.

For the other scalar state $|\Delta_s,l\rangle=|3,0\rangle$, with the choices in \eqref{case1_choiceab_action} we can raise the lowest weight state $|3,0\rangle$ to $|4,1\rangle$ and consistently lower it back to $|3,0\rangle$ as follows
\begin{subequations}
    \begin{align}
        &C_i|4,l_T=1\rangle=-\frac{2}{a}|3,0\rangle,\\
       |3,0\rangle:~~&T_i|3,0\rangle=3a|4,l_T=1\rangle,~~C_i|3,0\rangle=0.
    \end{align}
\end{subequations}

At this point there is nothing to stop us from raising the state $|4,1\rangle$ to $|5,\tilde{l}_T\rangle$ and higher up since the rest of $\mathcal{N}_C(3+n,\tilde{l}_T)$ with $n=1,2,3,\dots$ are well defined. For instance the next step in the ladder build from the $|3,0\rangle$ lowest weight state is
\begin{subequations}
    \begin{align}
        &C_i|5,\tilde{l}_T\rangle=-\frac{2}{a}|4,1\rangle,\\
       |4,1\rangle:~~&T_i|4,1\rangle=4a|5,\tilde{l}_T\rangle,~~C_i|4,1\rangle\neq0.
    \end{align}
\end{subequations}

Thus the choice of \eqref{case1_choiceab_action} leads to the following two towers of lowest weight states for the massless scalar on $dS_4$
\begin{subequations}\label{massless scalar towers}
    \begin{align}
        |0,0\rangle:&~~|0,0\rangle~\text{alone},\\
        |3,0\rangle:&~~|3+n,\tilde{l}\rangle=\frac{1}{\mathcal{N}_T(3,0)\dots\mathcal{N}_T(3+n,0)}\left(T_i\right)^n|3,0\rangle,~~n=1,2,3,\dots
    \end{align}
\end{subequations}
where the tower of $|0,0\rangle$ contains $|0,0\rangle$ alone while the second tower, the tower of $|3,0\rangle$ raises infinitely high with different values of spin at every step.

\section{Conclusions}
Looking at the de Sitter algebra, the SCT generators ($C_i$) act as lowering operators by lowering the scaling dimension $\Delta$, one unit at a time and Translation ($T_i$) generators act as raising operators. In this work we have identified these generators as the \emph{ladder operators}. In the context of raising and lowering one UIR to another, we noted that the action of the ladder operators are only meaningful for those states that correspond to discrete series representations. We set up our terminology based on how the scaling dimension changes under the action of the ladder operators, but along the way we also analyzed how the ladder operators effect spin. 

Defining the \emph{lowest weight state} as the state annihilated by the action of SCT generators we fixed the action of the ladder operators and investigated the implications of our choices for the UIRs that arise from the free massless scalar on $dS_4$. This meant considering $|0,0\rangle$ and $|3,0\rangle$ as the lowest weight sates. We saw that the lowest weight state $|0,0\rangle$ corresponds to a finite tower of a single step on its own while the lowest weight state $|3,0\rangle$ gives rise to an infinite ladder where each step consists of a UIR of with a different value of scaling dimension and spin.

We leave it for future work to broaden this analysis to also consider lowest weight towers build from lowest-weight states with nontrivial spin, to write down the generators in terms of annihilation and creation operators and explicitly show that their actions on field theoretic realizations in a complementary route and to compare with other ladder operators from literature such as those on the maximally compact subgroup label in the context of $dS_2$ \cite{Anninos:2023lin} and from the conformal algebra of $dS_4$ \cite{Letsios:2024snc}.

\begin{acknowledgments}

It's a pleasure to thank Carlo Iazeolla, Toshiyuki Kobayashi, Vasileios Letsios, Benjamin Pethybridge, Todor Popov, Joris Raeymaekers for stimulating discussions. The authors MB, EBG and GŞ acknowledge support from  TÜBİTAK BİDEB 2232B grant no 121C138 and GŞ also acknowledges support from
 TÜBİTAK ARDEB UPAG 1071 grant no 123N952 in the final stages of this work.
\end{acknowledgments}

\bibliographystyle{JHEP}

\begin{thebibliography}{99.}

\bibitem{Dobrev:1977qv}V. K. Dobrev, G. Mack, V. B. Petkova, S. G. Petrova and I. T. Todorov, \textit{Harmonic Analysis on the n-Dimensional Lorentz Group and Its Application to Conformal Quantum Field Theory}, Lect. Notes Phys. 63 (1977) 1.

\bibitem{sengor2021scalar} G. Sengor, C. Skordis, JHEP \textbf{02} (2024) 076

\bibitem{sun2021note} Z. Sun, Rev.Math.Phys. 37 (2025) 01, 2430007

\bibitem{Silva_note} A. Rios Fukelman, M. Sempé and G. A. Silva, JHEP \textbf{01} (2024) 011

\bibitem{Anninos:2023lin} D. Anninos, T. Anous, B. Pethybridge, G. Şengör, J.Phys.A \textbf{57} (2024) 2, 025401

\bibitem{Letsios:2024snc} V. A. Letsios, M. N. Semp{\'e}, G. A. Silva, Phys. Rev. D \textbf{111} (2025) 025018
\end{thebibliography}

\end{document}